\documentclass[12pt]{article}

 \usepackage{graphicx}
\usepackage{graphicx, epsfig, amssymb} 
\usepackage{bm}
\usepackage{amsfonts}
\usepackage{amsmath}
\usepackage{color}
\usepackage{amssymb}
\usepackage{graphicx}
\usepackage{epsfig}
\usepackage{amsfonts}

\date{\today}

\begin{document}
\title{\Large{\bf Shadows of Einstein-dilaton-Gauss-Bonnet black holes
} }

 \author{
{\large Pedro V. P. Cunha}$^{1,2}$, \
{\large Carlos A. R. Herdeiro}$^{1}$, \
{\large Burkhard Kleihaus}$^{3}$, \\
{\large Jutta Kunz}$^{3}$ and
{\large Eugen Radu}$^{1}$
\\
\\
$^{1}${\small Departamento de F\'\i sica da Universidade de Aveiro and  } \\ {\small  Centre for Research and Development  in Mathematics and Applications (CIDMA),
 } \\ {\small    Campus de Santiago, 3810-183 Aveiro, Portugal}
 \\ \texttt{\small  pintodacunha@tecnico.ulisboa.pt; herdeiro@ua.pt; eugen.radu@ua.pt}
 \\
 \\
$^{2}${\small CENTRA, Departamento de F\'\i sica, Instituto Superior T\'ecnico} \\ {\small Universidade de Lisboa, Avenida Rovisco Pais 1, 1049, Lisboa, Portugal}
 \\
 \\
$^{3}${\small Institute of Physics, University of Oldenburg, Oldenburg, 26111, Germany}
 \\ \texttt{\small  b.kleihaus@uni-oldenburg.de; jutta.kunz@uni-oldenburg.de}
}
\maketitle


\begin{abstract}
We study the shadows of the fully non-linear, asymptotically flat Einstein-dilaton-Gauss-Bonnet  (EdGB) black holes (BHs), for both static and rotating solutions.  We find that, in all cases, these shadows are \textit{smaller} than for  \textit{comparable} Kerr BHs, $i.e.$ with the same total mass and angular momentum {under similar observation conditions}.   In order to compare both cases we provide quantitative shadow parameters, observing in particular that the differences in the shadows mean radii are never larger than the percent level. Therefore, generically, EdGB BHs $cannot$ be excluded by 
(near future) shadow observations alone. On the theoretical side, we find no clear signature of some exotic features of EdGB BHs on the corresponding shadows, such as the regions of negative (Komar, say) energy density outside the horizon. We speculate that this is due to the fact that the Komar energy interior 
to the light rings (or more precisely, the surfaces of constant radial 
coordinate that intersect the light rings in the equatorial plane) is always smaller than the ADM mass, and consequently the corresponding shadows are smaller than those of comparable Kerr BHs. The analysis herein provides a clear example that it is the light ring impact parameter, rather than its ``size", that determines a BH shadow.  

\end{abstract}


\section{Introduction}
Ultraviolet theoretical inconsistencies of Einstein's General Relativity, such as its non-renormalizability~\cite{'tHooft:1974bx,Deser:1974cz,Deser:1974xq} and  the existence of singularities, have since long motivated the suggestion that higher curvature corrections should be taken into account, in an improved theory of gravity (see $e.g.$~\cite{Stelle:1976gc}). Inclusion of a finite set of such higher curvature corrections, however, generically leads to runaway modes (Ostrogradsky instabilities~\cite{Ostrogradsky:1850fid}) in the classical theory and a breakdown of unitarity due to ghosts, in the quantum theory. These undesirable properties can be simply diagnosed, at the level of the classic field equations, by the presence of third order time (and consequently also space, by covariance) derivatives. A natural way around this problem is to require a self-consistent model, obtained as a truncation of the higher curvature expansion, to yield a set of field equations without such higher order derivatives.

\bigskip

Lovelock~\cite{Lovelock:1971yv} first established, for vacuum gravity, what are the allowed curvature combinations so that the field equations have no higher than second order time derivatives. It turns out that, in a Lagrangian, these combinations are simply the Euler densities, particular scalar polynomial combinations of the curvature tensors of order $n$. Since the $n^{th}$ Euler density is a topological invariant in spacetime dimension $D=2n$ and yields a non-dynamical contribution to the action in dimensions $D\leqslant 2n$, an immediate corollary is that, in $D=4$ vacuum gravity, the most general Lovelock theory is a combination of the 0$^{th}$ and 1$^{st}$ Euler density, or in other words, General Relativity with a cosmological constant. The 2$^{nd}$ Euler density, known as the Gauss-Bonnet (GB) combination, is a topological invariant in $D=4$ and does not contribute to the dynamical equations of motion if included in the action. 

\bigskip

There is, however, a simple and natural way to make the GB combination dynamical in a $D=4$ theory:  couple it to a dynamical scalar field. This is actually a model that emerges naturally in string theory~\cite{Zwiebach:1985uq} (see also~\cite{Kanti:1995vq} for a discussion on this point), where the scalar field is the dilaton, and can be considered as a simple effective model to investigate the consequences of higher curvature corrections in $D=4$ gravity. The corresponding model takes the name of Einstein-dilaton-Gauss-Bonnet (EdGB) theory and is described by the action~\eqref{act} in section~\ref{section_model} below.

\bigskip

Black holes (BHs) in EdGB theory were first shown to exist, in spherical symmetry, by Kanti et al.~\cite{Kanti:1995vq}, wherein they were obtained numerically. These solutions, which moreover are perturbatively stable along their main branch~\cite{Kanti:1997br}, are asymptotically flat, regular on and outside an event horizon, and describe a horizon surrounded by a non-trivial dilaton profile. They circumvent some well-known no (real) scalar hair theorems, namely those by Bekenstein~\cite{Bekenstein:1972ny,Bekenstein:1995un} (see~\cite{Herdeiro:2015waa} for a recent review), due to the non minimal coupling of the dilaton to the geometry 
and the fact that if one associates some \textit{effective matter} with the GB term, then this represents \textit{exotic matter}, violating the typical energy conditions. One manifestation of this \textit{effective exotic matter} is that the BH solutions have regions of negative energy density outside the horizon. Another manifestation is that there is a minimal mass for BHs, determined by the GB coupling. We remark that the scalar hair of this BHs has no-independent conserved charge, thus being called~\textit{secondary}. See, $e.g.$~\cite{Torii:1996yi,Alexeev:1996vs,Melis:2005ji,Chen:2006ge,Chen:2008hk} for further discussions of these spherically symmetric solutions and some charged generalizations.\footnote{BH solutions of a closely related Horndeski model can be found in~\cite{Sotiriou:2013qea,Sotiriou:2014pfa}.}

\bigskip

Rotating BHs in EdGB theory were found, fully non-linearly in~\cite{Kleihaus:2011tg,Kleihaus:2015aje} (see also~\cite{Pani:2009wy,Pani:2011gy, Ayzenberg:2014aka, Maselli:2015tta} for perturbative studies). A minimal mass depending on the GB coupling still exists for these rotating solutions and, as a novel physical feature, some (small) violations of the Kerr bound in terms of ADM quantities are observed. Again, regions with negative energy density exist outside the horizon.

\bigskip

 In this paper, we shall investigate 
how the dGB term impacts on one particular observable feature of a BH: its shadow~\cite{Falcke:1999pj}.
 BH shadows can be roughly described as the silhouette produced by the BH when placed in front of a bright background. 
They are determined by the BH  absorption cross section for light 
at high frequencies.  Over the last few years there has been a renewed theoretical interest in this old concept, 
first discussed for the Kerr BH by Bardeen~\cite{Bardeen:1973tla}, due to observational attempts to measure the BH shadow of the supermassive BHs in our galactic center as well as that in the centre of M87~\cite{Lu:2014zja}. In particular, in~\cite{Cunha:2015yba,Vincent:2016sjq,Cunha:2016bjh}, the shadows of a type of hairy BHs that connect continuously to Kerr, within General Relativity and with matter obeying all energy conditions, called Kerr BHs with scalar hair~\cite{Herdeiro:2014goa,Herdeiro:2014ima,Herdeiro:2015gia}, have been studied. It has been pointed out that, generically, these shadows are smaller than those of a comparable Kerr BH, $i.e.$ a vacuum rotating BH with the same total mass and angular momentum. A possible interpretation of this qualitative behaviour is the following:  the total mass (and angular momentum) of the hairy BHs is now partly stored in the scalar field outside the horizon; in particular the existence of some energy outside the 
region of unstable spherical photon orbits, also referred to as photon region (see section~\ref{subsection_Ligh-rings}) \cite{Grenzebach:2014fha}, implies that less energy exists inside this region and hence the light rings should be smaller (within an appropriate measure) as compared to their vacuum counterparts and consequently so should be the shadows. 
 
\bigskip

The above interpretation raises an interesting question in relation to the BHs in EdGB theory. 
Since these have negative energy densities outside the horizon, 
how do these regions of \textit{effective exotic matter} impact on their shadows? 
In particular could there be a negative energy contribution 
outside the photon region that is sufficiently large to increase the shadow size with respect to a vacuum counterpart? 
We remark that for other non-vacuum solutions with physical matter, $i.e.$ obeying all energy conditions, the size of the shadow typically decreases with respect to the size of a comparable vacuum Kerr BH -- see $e.g.$~\cite{Takahashi:2005hy} for electrically charged BHs. 
However, larger shadows have also been observed, $e.g.$, in extended Chern-Simons gravity~\cite{Amarilla:2010zq} 
or brane world BHs~\cite{Amarilla:2011fx} which possess \textit{effective exotic matter}, similarly to EdGB.
Nevertheless, we shall see that for EdGB the shadows are always smaller with respect to the vacuum case, with the maximal deviation being of the order of only a few percent. For some work on BH shadows in different models see \cite{Cunha:2015yba,Grenzebach:2014fha,Amarilla:2010zq, Amarilla:2011fx, Tretyakova:2016ale, Abdolrahimi:2015rua, Abdolrahimi:2015kma, Shipley:2016omi, Abdujabbarov:2016hnw, Amir:2016cen, Johannsen:2015qca,Amarilla:2013sj}, and in particular~\cite{Younsi:2016azx} for perturbative EdGB BHs.

\bigskip

This paper is organized as follows. 
In Section~\ref{section_model} we describe the EdGB model and present its field equations. 
An overview of the known
BH solutions in this model, both static and stationary, 
is also provided there,
together with
the corresponding domain of existence and limiting cases. 
Then, in Section~\ref{section_shadows} we present the shadows for a representative sample of solutions 
and interpret the patterns obtained.
We close with a discussion in Section~\ref{section_discussion}.

\section{ The model and solutions}
\label{section_model}
 \subsection{ The field equations and general results }
 
We consider the Einstein-dilaton-Gauss-Bonnet (EdGB) model, described by the following action\footnote{In this work we shall use geometric units $c = G = 1$.}
\begin{eqnarray}  
S=\frac{1}{16 \pi  }\int d^4x \sqrt{-g} \left[R - \frac{1}{2}
 (\partial_\mu \phi)^2
 + \alpha  e^{-\gamma \phi} R^2_{\rm GB}   \right],
\label{act}
\end{eqnarray} 
where 
$\phi$ is the dilaton field, $\alpha $ is a parameter with units (length)$^2$ 
and
$R^2_{\rm GB} = R_{\mu\nu\rho\sigma} R^{\mu\nu\rho\sigma}
- 4 R_{\mu\nu} R^{\mu\nu} + R^2$ 
is the GB combination. 
Also, $\gamma$ is an input parameter of the theory\footnote{Since
the system possesses the symmetry $\gamma \to-\gamma$, $\phi\to -\phi$,
it is enough to consider strictly positive values of $\gamma$. Furthermore, in order to have a non-trivial coupling to the dilaton field, $\gamma\neq 0$.}, with most of the studies assuming $\gamma=1$. {Both $\gamma$ and $\phi$ are dimensionless.}

Varying the action~(\ref{act}) with respect to $g_{\mu\nu}$, we obtain\footnote{We follow the
conventions in Ref. \cite{Guo:2008hf}.  }
the  Einstein equations:
\begin{eqnarray}
\label{EGB-eq1}
 G_{\mu\nu}= T_{\mu\nu}^{\rm (eff)} \ ,
 \end{eqnarray} 
 where $G_{\mu\nu}$ is the standard Einstein tensor and the effective energy-momentum tensor reads
  \begin{eqnarray}
\label{Teff}
 T_{\mu\nu}^{\rm (eff)}\equiv \frac12\biggl[\nabla_{\mu}\phi\nabla_{\nu} \phi -\frac12 g_{\mu\nu}(\nabla\phi)^2\biggr]
-\alpha  e^{-\gamma\phi}T_{\mu\nu}^{\rm (GBd)} \ ,
 \end{eqnarray} 
 where the full expression for $T_{\mu\nu}^{\rm (GBd)}$ can be found in \cite{Kleihaus:2015aje}. Varying the action~(\ref{act}) with respect to the dilaton field, on the other hand, yields the scalar equation of motion, which reads:
\begin{eqnarray}
\label{dil-eq}
\Box  \phi =\alpha  \gamma e^{-\gamma\phi}  R^2_{\rm GB} \ .
\end{eqnarray} 

The EdGB model possesses BH and wormhole \cite{Kanti:2011jz} solutions, 
but no particle-like
solitonic configurations are known 
(for a review, see the recent work \cite{Blazquez-Salcedo:2016yka}),
although the coupling to matter leads, $e.g.$, to neutron stars
\cite{Pani:2011xm,Kleihaus:2016dui}.
Note that in contrast to the GR case, all  EdGB  solutions (with $\alpha\neq 0$)
 have been obtained numerically. 

In terms of the spherical-type coordinates 
$r,~\theta$ and $\varphi$, 
all known EdGB solutions possess at least two Killing vectors 
$\xi=\partial/\partial t$ and $ \zeta=\partial/\partial \varphi$
(where $t$ is the time coordinate).
Then a generic metric ansatz 
can be written as
\begin{eqnarray}
\label{metric}
ds^2=g_{rr}dr^2+g_{\theta\theta}d\theta^2+g_{\varphi\varphi}d\varphi^2+2 g_{\varphi t} d\varphi dt+g_{tt}dt^2~,
\end{eqnarray}
where 
$g_{\mu \nu}$ and the scalar $\phi$
are functions of $(r,\theta)$. Moreover, we can set $\phi(\infty)=0$ without any loss of generality {(any other choice would correspond to a rescaling of the radial coordinate in~\eqref{metric} \cite{Kleihaus:2015aje})}. 
The ADM (Arnowitt-Deser-Misner)  mass $M$ and angular momentum $J$ 
are read off, as usual, from the asymptotic expansion 
\begin{eqnarray}
\label{asym}
g_{tt} =-1+\frac{2M}{r}+\dots,~~g_{\varphi t}=-\frac{2J}{r}\sin^2\theta+\dots~.~~
\end{eqnarray}

One can also define a global dilaton measure $D$ from the asymptotic expansion 
of the scalar field, $\phi=-D/r+\dots$ which however is not an independent quantity, 
since the dilaton field does not qualify as primary hair 
\cite{Kanti:1995vq}, \cite{Kleihaus:2015aje}.

\subsection{The static EdGB black holes}

Consider for the moment the static, spherically symmetric solutions ($J=0$). 
Close to the event horizon, these solutions possess an approximate expression  
as a power series in $r-r_H$, where 
$r_H$
 is the radial coordinate of the horizon. 
{ In particular, in Schwarzschild
coordinates one finds
$\phi(r)=\phi_H+\phi_1(r-r_H)+\dots$, where $\phi_1$ satisfies a quadratic equation
(see $e.g.$ \cite{Kanti:1995vq}, 
\cite{Torii:1996yi},
\cite{Alexeev:1996vs},
\cite{Kleihaus:2015aje}).
Since the scalar field is real,} the discriminant of the quadratic equation is required to be positive, yielding the condition:
 \begin{eqnarray}
 \label{cond1}
 1-96\alpha^2\gamma^2\frac{e^{-2\gamma \phi_H}}{A_H^2/(16\pi^2)} \geq 0,
 \end{eqnarray}  
where $A_H$ is the event horizon area. Eventually, this condition will be violated after some \textit{limiting solution} is reached, beyond which solutions cease to exist in the parameter space. 
For a given $\gamma$, all solutions can be obtained continuously in the parameter space. 
When appropriately scaled they
form a line, starting from the smooth GR limit ($\phi\to 0$ as $\alpha \to 0$), and ending at the limiting solution. 
The existence of the latter places a lower bound on the BH horizon radius. 
It actually also implies the existence of a lower bound on the
BH mass. In particular, as discussed in \cite{Kanti:1995vq,Pani:2009wy}, the static EdGB solutions with $\gamma=1$ are limited to the parameter range $0\leq\alpha/M^2\lesssim 0.1728$. A rather similar behaviour holds\footnote{Note that solutions seem to exist for any nonzero value of $\gamma$.} for  $\gamma\neq 1$. 

\begin{figure}[h!]
\begin{center}   
\hspace*{0.7cm}\hspace*{-1.6cm}\includegraphics[width=9.cm]{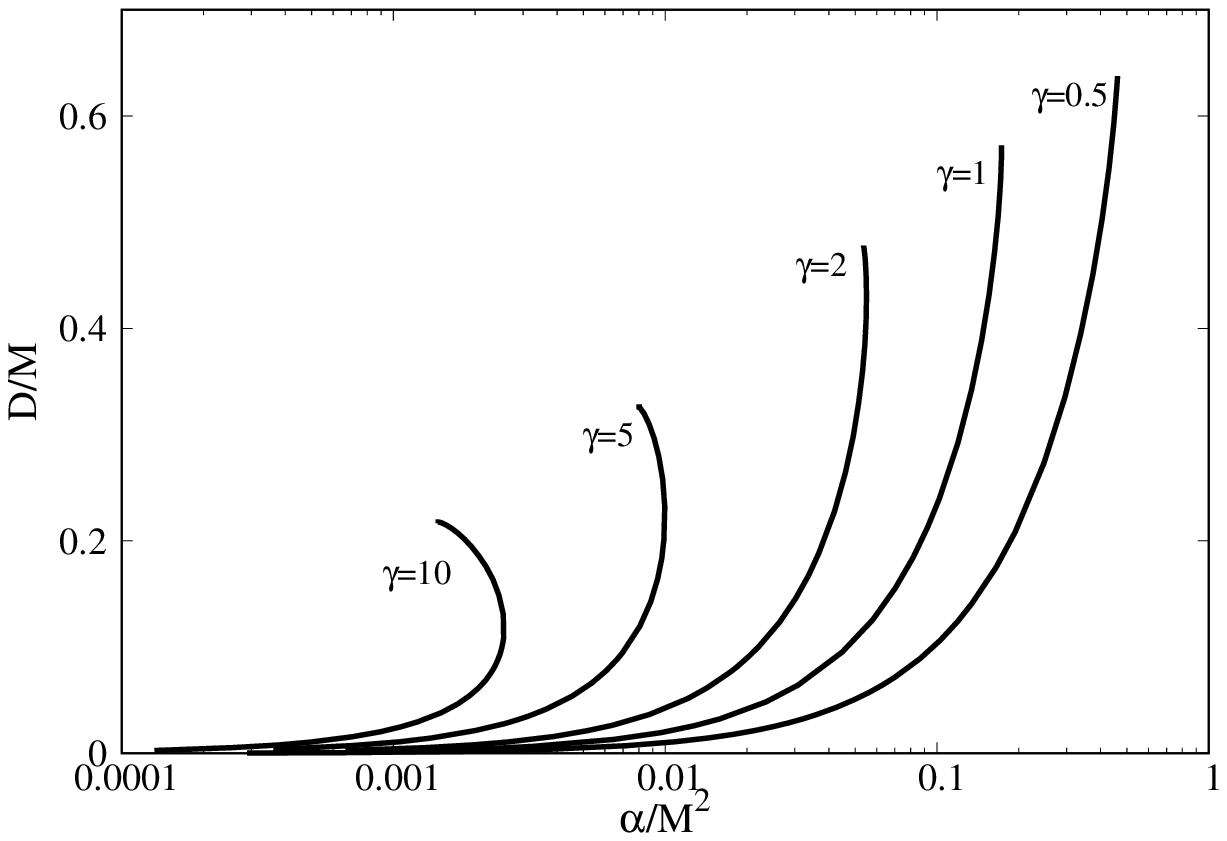}\includegraphics[width=9.cm]{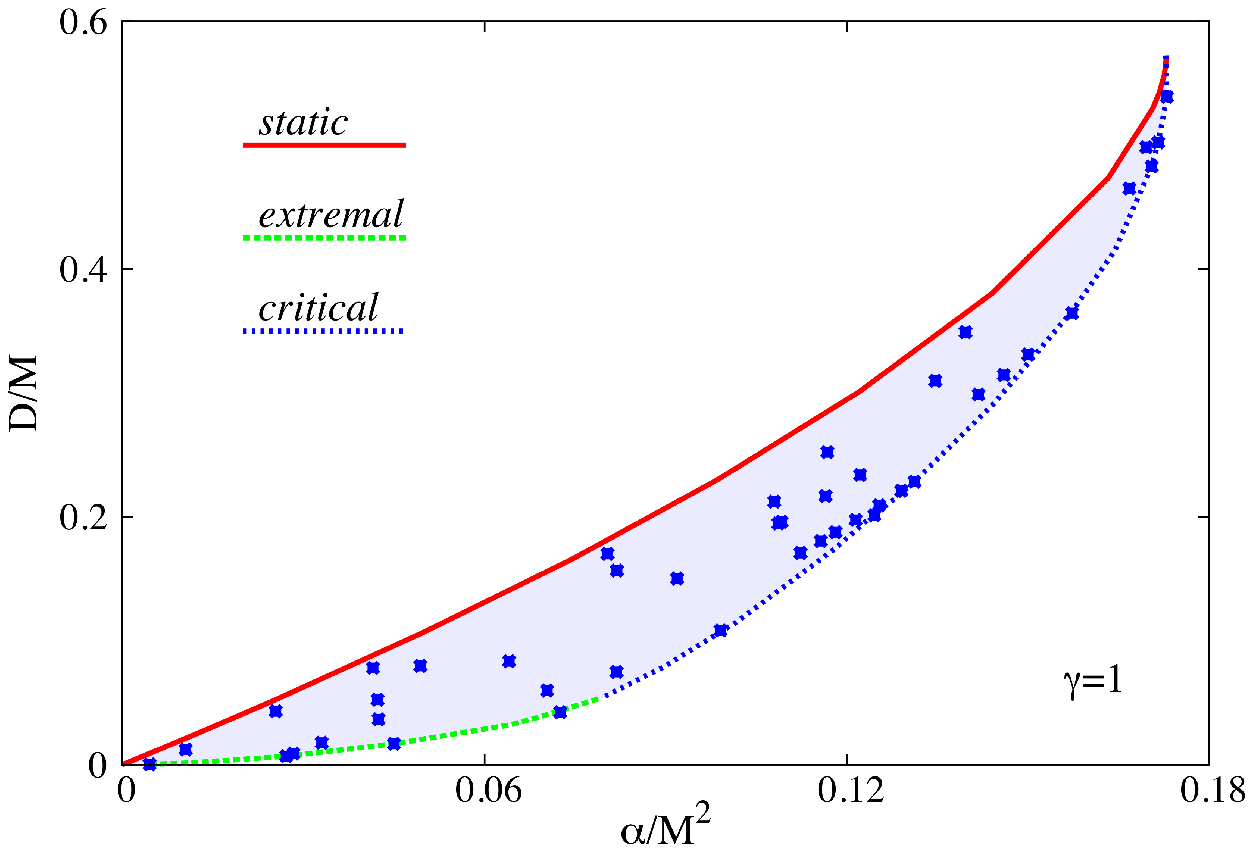}
\put(-409.5,47){{\circle{3}}}
\put(-423,68.5){{\circle{3}}}
\put(-409.5,45){  {$b$}}
\put(-423,68.5){  {$a$}}
\end{center}
  \vspace{-0.5cm}
\caption{ (Left) Domain of existence of static EdGB BHs in a $D/M$ vs.~$\alpha/M^2$ diagram with several values of $\gamma$. {The points $a$ and $b$ depict the limiting and critical solutions respectively for $\gamma=10$.} (Right) Domain of existence of spinning solutions with $\gamma=1$. The set of considered (spinning) solutions in Fig.~\ref{fig-r} and Fig.~\ref{fig-sig} are shown here as highlighted points.
}
\label{DM}
\end{figure}

Solutions no longer exist if the ratio $\alpha/M^2$ is larger than a critical value, which decreases with increasing $\gamma$. The configuration at this maximal value is dubbed the \textit{critical solution}, which needs not to coincide with the limiting solution. In particular, for large enough $\gamma$, the solution line can be extended backwards in $\alpha/M^2$, into a ``secondary branch'', after the critical configuration is reached \cite{Guo:2008hf}; this secondary branch eventually terminates at the limiting solution.
Some of these features can be seen in an $(\alpha,D)$-diagram of solutions with different $\gamma$, as shown in Figure  \ref{DM} (left). In particular, notice how for sufficiently large $\gamma$ values 
it is possible to have two different values of $D/M$ for the same $\alpha/M^2$, which indicates the presence of two branches.
According to arguments from
catastrophy theory, the stability should change at the critical solution, so
that the solutions along the secondary branch will be unstable \cite{Torii:1996yi}.

 \subsection{The spinning EdGB black holes}
{Spherically symmetric BHs typically possess spinning
generalizations}.
However, so far only the $\gamma=1$
case has been explored in the literature.
These BHs
 were first obtained at the 
fully non-linear  level
in~\cite{Kleihaus:2011tg}
(see also~\cite{Pani:2009wy,Pani:2011gy, Ayzenberg:2014aka, Maselli:2015tta} 
for perturbative results). 
 Similar to the GR case,
these BHs possess a $\mathbb{Z}_2$ symmetry along the equatorial plane ($\theta=\pi/2$) 
and are obtained by solving  the field equations~\eqref{EGB-eq1} and~\eqref{dil-eq} 
subject to appropriate boundary conditions that are detailed in~\cite{Kleihaus:2015aje}.

The domain of existence of EdGB  BHs is bounded by four sets of solutions:  
$i)$ the set of static ($i.e.$ spherically symmetric) EdGB BHs with $J=0$;
$ii)$ the set of extremal ($i.e.$, zero temperature) EdGB BHs;
$iii)$ the set of critical solutions;
 and $iv)$ the set of GR solutions -- the Kerr/Schwarzschild BHs with $\alpha=0$. In Fig.~\ref{fig-r} and Fig.~\ref{fig-sig} the boundary line displayed includes the sets $ii)$ and $iii)$. 

The general critical solutions are the rotating generalization of the static case, while the extremal set does not appear to be regular on the horizon.\footnote{Perturbing the extremal vacuum solution in $\alpha$, the scalar field/metric develops singularities at the poles in first/second order in $\alpha$.}
Moreover, the mass of the EdGB rotating BHs is always bounded from below, whereas the angular momentum can (slightly) exceed the Kerr bound, which is given by {$J \leqslant M^2$}.
Further details on these aspects together with various plots of the domain of existence are found in 
\cite{Kleihaus:2015aje}.
Here we give the domain of existence in $(\alpha,D)$-variables [Figure 1 (right)]
and in  $(\alpha,J)$-variables
 (Figure  \ref{fig-r}).

\section{Shadows}
\label{section_shadows}

\subsection{Light rings}
\label{subsection_Ligh-rings}

As it is well described in the literature, the Kerr spacetime supports unstable photon orbits with a fixed {Boyer-Lindquist} radial coordinate, $i.e.$, the photon region
\cite{Bardeen:1973tla}. A subset of the latter is restricted to the equatorial plane $(\theta=\pi/2)$, and comprises two independent circular photon orbits with opposite rotation senses, dubbed here as \textit{light rings}.  Such orbits are not unique to the Kerr spacetime and have an intrinsic relation to the BH shadow. In particular, unstable light rings embody a threshold of stability between equatorial null geodesics that scatter to infinity and ones that plunge into the BH. Consequently, light rings account for the shadow edge in observations restricted to the equatorial plane's line of sight (provided both exist).   
Following \cite{Cunha:2016bjh}, the light ring positions can be obtained by analysing the following condition in the equatorial plane:
\begin{equation}\partial_rh_\pm=0,\quad \textrm{with}\quad h_\pm=\frac{-g_{t\varphi}\pm\sqrt{g^2_{t\varphi}-g_{tt}g_{\varphi\varphi}}}{g_{tt}}.\end{equation}
Recalling the Kerr case, each sign $\pm$ leads to one of the two light rings. Curiously, although the EdGB BHs discussed in this paper are fully non-linear solutions (rather than perturbations of Kerr), the light ring qualitative structure still appears to be the same as in Kerr. However, notice that for other families of solutions this is not always the case. For instance, multiple light rings can appear for BHs with scalar hair, some of which are stable \cite{Cunha:2016bjh}.\\

\subsection{Characterizing the shadow}

Assuming that a suitable light source is present to provide contrast, a BH casts a black region in an observer's sky, commonly called the BH shadow. Although some characteristics are observer dependent~\cite{Vincent:2016sjq}, the size and shape of the shadow are essentially a manifestation of the spacetime properties close to the BH, depending for instance on the light ring characteristics. Consequently, instructive physics can be inferred from such observations. \\

\begin{figure}[ht]
\begin{center}
\includegraphics[width=6cm]{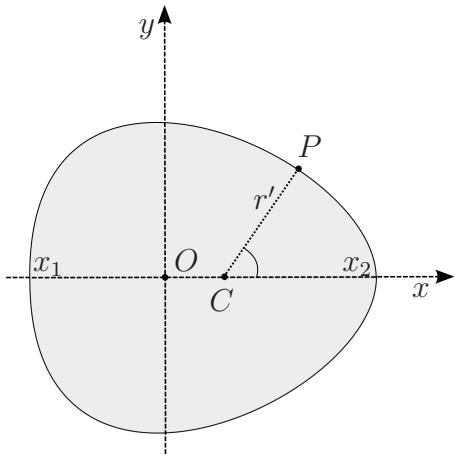}
\put(-60,113.3){$P$}
\put(-76.7,93.3){$r'$}
\put(-120,160){$y$}
\put(-16.7,60){$x$}
\put(-93.3,55){$C$}
\put(-106.7,70){$O$}
\put(-43,70){$x_2$}
\put(-160,70){$x_1$}
\end{center}
\caption{\small Representation of a BH shadow in the $(x,y)$ image plane of the observer.
}
\label{fig-shadow}
\end{figure}

Consider the dummy shadow in Fig. \ref{fig-shadow}, represented in the image plane of the observer. 
A Cartesian parametrization $(x,y)$ is used, where the $x$-axis is defined to be parallel to the azimuthal Killing vector $\zeta=  \partial/\partial\varphi$ at the observer's position. The origin $(0,0)$ of this coordinate system, defined as point $O$ in Fig. \ref{fig-shadow}, corresponds to the direction pointing towards the center of the BH $-\partial/\partial r$ (from the reader into the paper). 

The point $C$ in the figure, taken to be the center of the shadow, is such that its abscissa is given by $x_C=(x_{\textrm{max}}+x_{\textrm{min}})/2$, where $x_{\textrm{max}}$ and $x_{\textrm{min}}$ are respectively the maximum and minimum abscissae of the shadow's edge. If the observer is in the equatorial plane ($\theta=\pi/2$), which will be assumed throughout the paper, then the shadow inherits along the $x$-axis the spacetime reflection symmetry, giving $y_C=0$. Since the points $C$ and $O$ need not to coincide, a specific feature of a shadow is the {displacement} $x_C$ between the shadow and the center of the image plane $O$.

A generic point $P$ on the shadow's edge is at a distance $r'$ from $C$, which is defined as $r'\equiv \sqrt{{y_P}^2 + {(x_P-x_C)}^2}$. 
Given the line element $ds^2=dx^2 + dy^2$, the perimeter $\mathcal{P}$ of the shadow, its {average radius} $\bar{r}$ and the {deviation from sphericity} $\sigma_r$ are defined by:
\begin{equation}\oint ds \equiv \mathcal{P},\qquad \bar{r}\equiv \frac{1}{\mathcal{P}}\oint r'\,ds,\qquad \sigma_r=\sqrt{\frac{1}{\mathcal{P}}\oint {\left(1-\frac{r'}{\bar{r}}\right)}^2\,ds}.\end{equation}
All these parameters are expressed in units of the ADM mass $M$.\\
In some cases, it is possible to compare the shadow parameters of a given EdGB solution with the ones from a Kerr BH with the same ADM mass $M$ and angular momentum $J$. Hence, let us also define the relative deviations to the Kerr case\footnote{An analytical expression for the Kerr shadow, as seen by an observer with zero angular momentum (ZAMO), can be found in \cite{Cunha:2016bpi}.}:
\begin{equation}\delta_r =\frac{\bar{r}-\bar{r}_\textrm{kerr}}{\bar{r}_\textrm{kerr}},\qquad\delta_\sigma =\frac{\sigma_r-\sigma_\textrm{kerr}}{\sigma_\textrm{kerr}},\qquad \delta_{x_C}=\frac{ x_C -{x_C}_\textrm{kerr} }{{x_C}_\textrm{kerr}}.\end{equation}

\subsection{Rotating EdGB BHs}

Due to the existence of a hidden constant of motion - the Carter constant - the edge of the Kerr shadow can be obtained in a closed analytical form \cite{Bardeen:1973tla,Grenzebach:2014fha,Cunha:2016bpi}. However, EdGB BHs are not expected to have such a property, since they all appear to be of Petrov type I \cite{Kleihaus:2015aje}. This is consistent with the perturbative results in \cite{Ayzenberg:2014aka}. As a consequence, in general the shadow of the latter has to be obtained numerically through the standard \textit{backwards ray-tracing} framework \cite{Johannsen:2015qca,Psaltis:2010ww}. In order to generate a virtual image of the shadow, this method requires propagating null geodesics ``backwards in time'', where a high frequency approximation is assumed, starting from the observer's position and determining the source of each light ray. Different points in the image plane correspond to different directions in the observer's sky, and hence to different initial conditions of the geodesic equations. The shadow is precisely the set of all those initial conditions which induce geodesics with endpoints on the event horizon, when propagated backwards in time. Since the event horizon is not a source of any light (classically), the shadow actually embodies a lack of radiation\footnote{We are implicitly assuming that there is no glowing matter in front of the BH.}.\\

The geodesic propagation method described above is necessary to compute most of the shadow edge. However, the points $x_1$ and $x_2$ in Fig. \ref{fig-shadow}, where the edge intersects the $x$-axis, can be computed using a highly precise local method. In particular, for an observer in the equatorial plane, light rings are the orbits responsible for these intersection points. The impact parameter $\eta=L/E$ will play here a crucial role, where $E$ and $L$ are respectively the photon's energy and axial angular momentum with respect to a static observer at infinity. Moreover, these quantities are constants of geodesic motion, connected to the Killing vectors of the spacetime $\xi=\partial/\partial t$ and $\zeta=\partial/\partial \varphi$.
The function $h_\pm$ will now be helpful again, as the value of $\eta$ in a given light ring orbit is provided simply by $\eta=h_\pm$, computed at that position \cite{Cunha:2016bjh}.

The precise relation between the image coordinate $x$ and the impact parameter $\eta$ depends on the choice for the observer's frame, but also on how $x$ is constructed in terms of observation angles. Following \cite{Cunha:2016bjh,Cunha:2016bpi}, the $x$ coordinate is defined to be directly proportional to an observation angle $\beta$ along that axis: $x=-\tilde{R}\,\beta,$ where the perimetral radius $\tilde{R}\equiv \sqrt{g_{\varphi\varphi}}$ is computed at the observer's position. By computing the projection of the photon's 4-momentum onto a ZAMO frame \cite{Cunha:2016bjh,Cunha:2016bpi}, the relation $\sin\beta = \eta/(A_0+\eta\,B_0)$ can be derived (if $y=0$), where the following quantities are computed at the position of the observer: $A_0=g_{\varphi\varphi}/\sqrt{D},\quad B_0=g_{t\varphi}/\sqrt{D}$, with $D\equiv g_{t\varphi}^2-g_{tt}g_{\varphi\varphi}$. This leads to the relation (with $y=0$):
\begin{equation}x=-\tilde{R}\arcsin\left(\frac{\eta}{A_0+\eta B_0}\right).\end{equation}
For the sake of the argument, consider also a very far away observer ($r\to \infty$). In these conditions we obtain the very simple relation $x=-\eta$.
By computing $\eta_1$ and $\eta_2$ for each of the two light rings, we can obtain the shadow radius $\bar{r}_x$ on the $x$-axis simply with $\bar{r}_x=|x_1-x_2|/2$, where each $x$ is evaluated from the respective $\eta$. Notice that this is a local method, in the sense that it does not require the evolution of a geodesic throughout the spacetime. Hence, obtaining a very precise $\bar{r}_x$ value only depends on knowing $\eta$ at the light rings with sufficiently high accuracy. Furthermore, by comparing this $\bar{r}_x$ value with the one obtained with ray-tracing, we can estimate that the precision of the latter is around $\sim 0.08\%$.\\

\begin{figure}[ht]
\begin{center}
\hspace*{-1.6cm}\includegraphics[width=9.8cm]{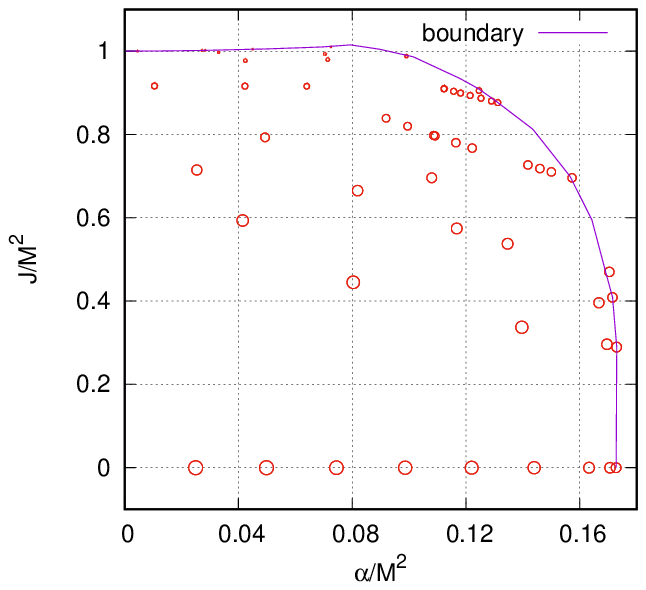}\includegraphics[width=9.8cm]{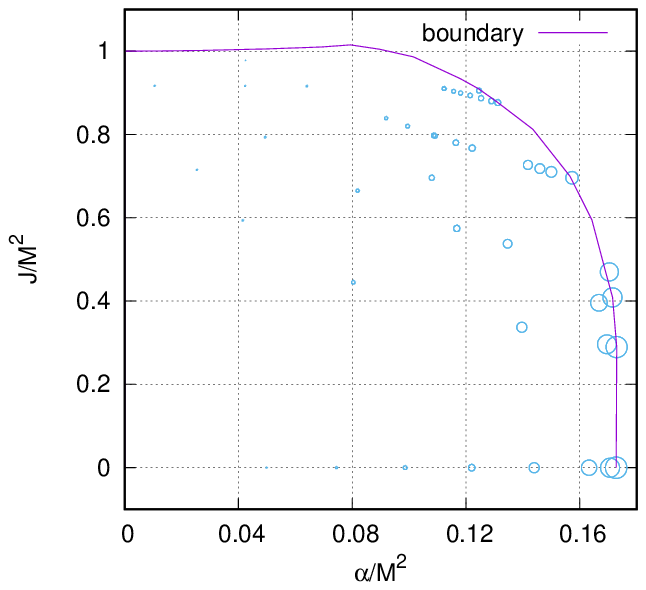}
\put(-425,260){{$\bar{r}-4.68M$}}
\put(-130,260){{$\delta_r$}}
\put(-500,65){$\simeq$ 0.24}
\put(-503,207){$\simeq$ 0.1}
\put(-58,168){$\simeq-0.9$}
\put(-58,46){$\simeq-1.5$}
\put(-205,205){$\simeq-0.08$}
\end{center}
\caption{\small Representation of $(\bar{r}-4.68M)$ (left) and $\delta_r$ (right) for EdGB solutions with $\gamma=1$, in a $\alpha/M^2$ vs. $J/M^2$ diagram. Each circle {radius} is proportional to the quantity represented, {with some values also included for reference}. All the values of $\delta_r$ are negative, with the maximum deviation to Kerr being around $\simeq -1.5\%$.
}
\label{fig-r}
\end{figure}

\begin{figure}[ht]
\begin{center}
\hspace*{-1.6cm}\includegraphics[width=9.8cm]{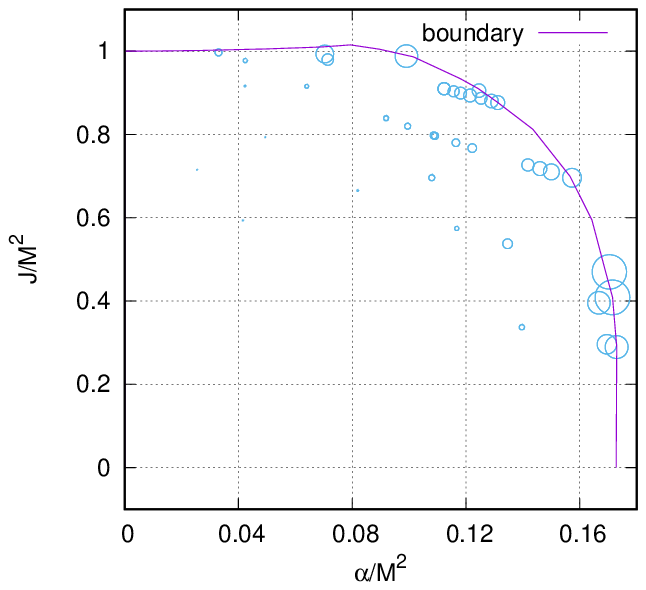}\includegraphics[width=9.8cm]{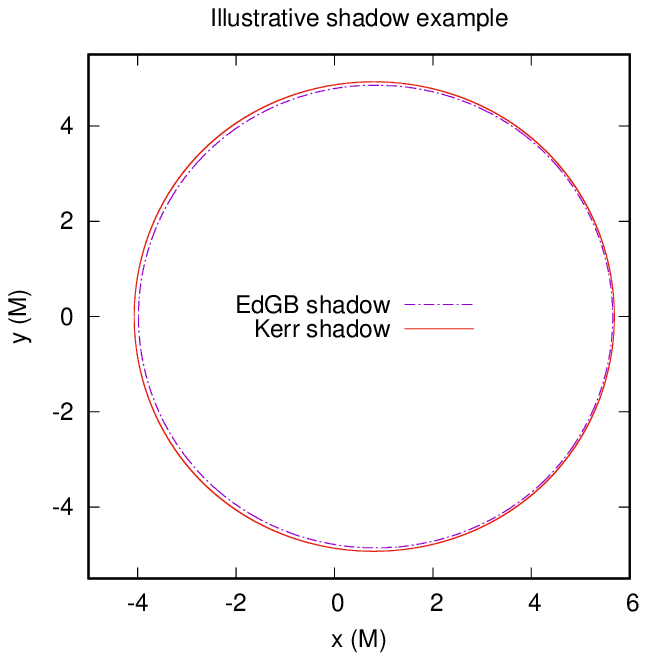}
\put(-410,260){{$|\delta_\sigma|$}}
\put(-355,217){$\sim 7$}
\put(-430,165){$\sim 1$}
\end{center}
\caption{\small (Left) Representation of $|\delta_\sigma|$ for EdGB solutions with $\gamma=1$, in a $\alpha/M^2$ vs. $J/M^2$ diagram. Each circle {radius} is proportional to the quantity represented, {with some values also included for reference}. All the values of $\delta_\sigma$ are negative. 
(Right) Depiction of the shadow edge of a EdGB BH with $\gamma=1$ and $(\alpha/M^2,J/M^2)\simeq(0.172,0.41)$, yielding $\bar{r}\simeq 4.85$, $\sigma=0.3$, $x_C=0.84$; the radial deviation $\delta_r$ with respect to the comparable Kerr case is $\simeq -1.35\%$. {The observer is at a perimetral radius $15M$.}
}
\label{fig-sig}
\end{figure}

The data of the EdGB shadows, computed with ray-tracing, is represented in Fig. \ref{fig-r} and Fig. \ref{fig-sig}, where a dilaton coupling $\gamma=1$ is assumed. The observer is always placed in the equatorial plane, at a radial coordinate such that $\tilde{R}=\sqrt{g_{\varphi\varphi}}=15M$.\\

In the left of Fig. \ref{fig-r}, the size of each circle represents the value of the shadow radius $\bar{r}$ for several EdGB solutions. In order to make the differences across the solution space more apparent, the circle {radius} is proportional to $\bar{r}-4.68M$. In other words, a vanishing circle (in this plot only) represents $\bar{r}=4.68M$. With this depiction, it is clear that - as a rule of thumb - increasing either $J$ or $\alpha$ decreases the shadow size. However, from an observational\footnote{{For a given BH under observation, the quantities $M$, $J$ and $\tilde{R}$ are all assumed to be known.}} point of view, it is much more relevant to compare the shadow prediction of an EdGB model with the one of a comparable\footnote{{The shadows are comparable if $M$, $J$ and the observation distance $\tilde{R}=\sqrt{g_{\varphi\varphi}}$ are the same.}} Kerr BH with the same $M$ and $J$. 
In particular, on the right of Fig. \ref{fig-r} the relative differences of the shadow size $\delta_r$ with respect to Kerr is represented in a circle plot. All deviations are negative, with the largest ones (in absolute) around $\simeq -1.5\%$. As (another) rule of thumb, increasing $\alpha/M^2$ appears to lead to larger radial deviations from Kerr. In particular, the spherically symmetric EdGB line ($J=0$) includes some of the largest $|\delta_r|$ values. As a side note, the data represented by the smallest circles in the right of Fig. \ref{fig-r} correspond to deviations around $\sim 0.08\%$, which is about the estimated numerical accuracy. \\

For completeness, the deviations\footnote{{Additional measures of EdGB shadow shapes are possible, but they resemble closely Kerr ones.}} of $\sigma_r$ with respect to Kerr are represented in the left of Fig. \ref{fig-sig}. Curiously, all values of $\delta_\sigma$ are negative, which means that EdGB shadows are more ``circular'' than the corresponding Kerr case. Hence, the GB term appears to soften the spin deformations that exist on the Kerr shadows. Moreover, notice how the largest $|\delta_\sigma|$ values can be found close to the critical boundary in solution space.
Additionally, the deviations $\delta_{x_C}$ can be both positive and negative, although a plot for this quantity is not shown.

In order to display an illustrative shadow case, in the right of Fig.~\ref{fig-sig} we have the representation of a EdGB shadow edge in the image plane, together with the comparable Kerr one. Although the difference between the curves is barely visible, amounting to a variation of only $\simeq-1.35\%$ in the shadow size, the case here depicted has one of the largest values of $|\delta_r|$ for $\gamma=1$. Such an example reinforces the idea that shadow observations are very unlikely to constrain EdGB BH models in the near future.

\subsection{Static EdGB BHs}

Until this point we discussed only the shadows of EdGB solutions for dilatonic coupling $\gamma=1$. Repeating the above analysis for other values of $\gamma$ would be rather cumbersome. Nevertheless, as discussed in the previous subsection, some of the largest $\bar{r}$ deviations occur within the static case. Therefore this can be considered as an incentive to explore other values of $\gamma$, while restricting ourselves to $J=0$. This will provide some insight on the effect of the $\gamma$ parameter without much more effort. \\

For the static case ($J=0$) the shadow is a circle due to the spherical symmetry of the spacetime. Using this property, we have $\bar{r}=\bar{r}_x$, which allows us to use the high precision method described before, thus obtaining the shadow edge without having to resort to any ray-tracing. Notice that in this case $\sigma_r$ and $x_C$ are both zero due to the spherical symmetry.\\

\begin{figure}[t]
\begin{center}
\includegraphics[width=11cm]{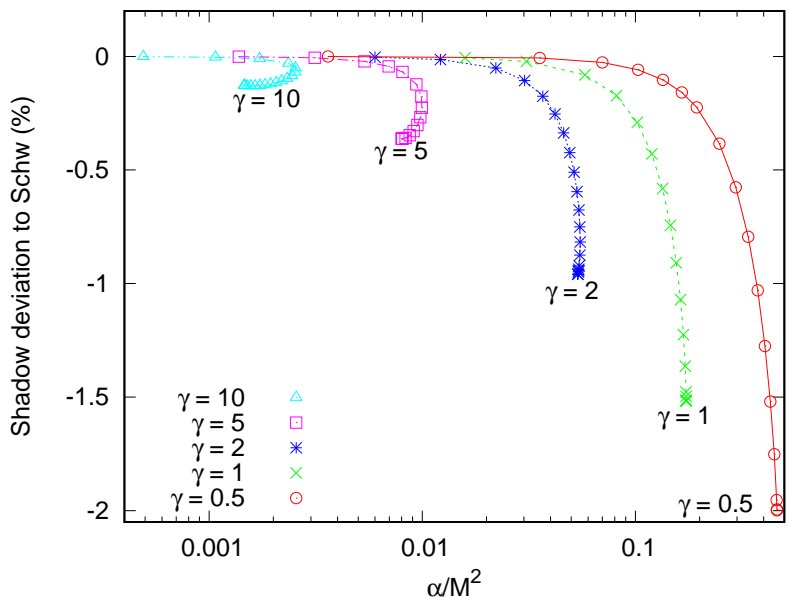}
\end{center}
\caption{\small  Representation of $\delta_r$ for static EdGB BHs, computed with respect to the Schwarzschild case. Data for different $\gamma$ values is displayed as a function of $\alpha/M^2$. All deviations are negative. {The displayed lines only interpolate the numerical data, with colors red, green, blue, pink and light blue respectively for $\gamma=\{0.5,1,2,5,10\}$. The observer's perimetral radius was set at $15M$.}
}
\label{fig-gamma}
\end{figure}

The radial deviations $\delta_r$ of static EdGB shadows with respect to those of a comparable Schwarzschild BH are represented in Fig. \ref{fig-gamma}, for different $\gamma$ values.
The data suggests a scenario 
where for a fixed value of $\alpha/M^2$ the deviations on the stable branches are larger if we increase $\gamma$; however, after entering the domain of the secondary (unstable) branches, $\gamma$ has to decrease in order to yield larger deviations. Furthermore, for a given $\gamma$, the maximum deviation always appears to occur at the limiting solution, with this maximal deviation being larger for smaller $\gamma$ values. For instance,
$\gamma=0.5$ can lead to shadows $\simeq 2\%$ smaller than for Schwarzschild, whereas for $\gamma=1$ all deviations are below $1.5\%$. 

\section{Discussion}
\label{section_discussion}

The shadow of a EdGB BH is always smaller than the comparable Kerr one. However, the deviations observed are always smaller (in modulus) than a few percent ($\sim 1\%$). Since such differences are below the expected resolution of planned observations {($\sim 6\%$ as anticipated in }\cite{Johannsen:2016vqy}), it is unlikely that in the near future any shadow measurement can exclude or restrict EdGB models. {Nevertheless, the present study was not exhaustive; it leaves, for instance, studies for different inclinations and distances as future work.}\\
Since EdGB theory possesses unusual features such as \textit{effective exotic matter}, it might come as a surprise that there are no significant effects at the level of the shadow. 
However, this \textit{effective exotic matter} is concentrated 
close to the horizon, such that there is no negative energy contribution
outside the photon region that could significantly affect the shadow's size.
At the same time any near-horizon odd effects are concealed from a remote observer by the shadow.\\
It may come as another surprise, that
the light ring size\footnote{The perimetral radius $\sqrt{g_{\varphi\varphi}}$ in $M$ units can be used as an invariant measure for the light ring size.} 
of EdGB BHs can, for instance, change by as much as $\simeq 4\%$, when considering the static case with $\gamma=0.5$,
and this effect will increase with further decreasing $\gamma$.
The natural question is then: why are the deviations in the shadow size not larger? For the sake of the argument consider the static case, where it becomes clear that the critical ingredient for the shadow radius is the impact parameter $\eta$, and not the light ring size. Naturally, there is a strong correlation between both concepts, but at the end of the day what matters is the value of the impact parameter. We would like to point out that this observation is often not clear enough in the literature: a large variation of the light ring size does not have to lead to equally large variations of the shadow radius.

\section*{Acknowledgments}

P.C. is supported by Grant No.  PD/BD/114071/2015 under the FCT-IDPASC Portugal Ph.D. program and by the Calouste Gulbenkian Foundation under the Stimulus for Research Program 2015.
C.H. and E. R. acknowledge funding from the FCT-IF programme.  This project has received funding
from  the  European  Union's  Horizon  2020  research  and  innovation  programme  under  the  Marie
Sklodowska-Curie grant agreement No 690904 and by the CIDMA project UID/MAT/04106/2013.
B.K. and J.K. gratefully acknowledge discussions with
Arne Grenzebach, Claus L\"ammerzahl and Volker Perlick,
as well as support by the DFG
Research Training Group 1620 ``Models of Gravity''
and by the grant FP7, Marie Curie Actions, People,
International Research Staff Exchange Scheme (IRSES-606096).

 \begin{small}
 
 \end{small}

 \end{document}